\documentclass[%
  reprint,
 amsmath,amssymb,
 aps,
prb,
]{revtex4-1}
\usepackage{graphicx}
\usepackage{dcolumn}
\usepackage{bm}
\usepackage{amsmath}
\usepackage[super]{nth}

\RequirePackage{hyperref}
\hypersetup{%
    linktocpage,
    bookmarksdepth = section,
    colorlinks = false,
    hidelinks = true
}
\RequirePackage[capitalise,nameinlink]{cleveref}

\begin{document}

\title{Impact of Substrate on Tip-enhanced Raman Spectroscopy --- A Comparison of Frequency Domain Simulations and Graphene Measurements}

\author{Hudson Miranda$^1$}
\author{Cassiano Rabelo$^1$}
\author{Luiz Gustavo Can\c{c}ado$^2$}
\author{Thiago L. Vasconcelos$^3$}
\author{Bruno S. Oliveira$^3$}
\author{Florian Schulz$^4$}
\author{Holger Lange$^{4,5}$}
\author{Stephanie Reich$^6$}
\author{Patryk Kusch$^6$}
\author{Ado Jorio$^{1,2}$}
\affiliation{$^1$Graduate Program in Electrical Engineering - UFMG - CEP: 31270-901, Belo Horizonte, MG, Brazil}
\affiliation{$^2$Departamento de F\'\i sica  - UFMG - CEP: 31270-901, Belo Horizonte, MG, Brazil}
\affiliation{$^3$Divis\~{a}o de Metrologia de Materiais - Inmetro - CEP: 25250-020, Duque de Caxias, RJ, Brazil}
\affiliation{$^4$Institute for Physical Chemistry, University of Hamburg, Martin-Luther-King Platz 6, 20146 Hamburg, Germany}
\affiliation{$^5$The Hamburg Centre for Ultrafast Imaging, 22761 Hamburg, Germany}
\affiliation{$^6$Department of Physics, Freie Universit\"at Berlin, Arnimallee 14, D-14195 Berlin, Germany}

\date{\today}

\begin{abstract}
Tip-enhanced Raman spectroscopy (TERS) has reached nanometer spatial resolution for measurements performed at ambient conditions and sub-nanometer resolution at ultra high vacuum. Super-resolution (beyond the tip apex diameter) TERS has been obtained, mostly in the gap mode configuration, where a conductive substrate localizes the electric fields. Here we present experimental and theoretical TERS to explore the field distribution responsible for spectral enhancement. We use gold tips of $40\pm 10 \ \text{nm}$ apex diameter to measure TERS on graphene, a spatially delocalized two-dimensional sample, sitting on different substrates: (i) glass, (ii) a thin layer of gold and (iii) a surface covered with $12\ \text{nm}$ diameter gold spheres, for which $6\ \text{nm}$ resolution is achieved at ambient conditions. The super-resolution is due to the field configuration resulting from the coupled tip-sample-substrate system, exhibiting a non-trivial spatial surface distribution. The field distribution and the symmetry selection rules are different for non-gap vs. gap mode configurations. This influences the overall enhancement which depends on the Raman mode symmetry and substrate structure.
\end{abstract}

\maketitle

\section{Introduction}
\label{s:intro}

Tip-enhanced Raman spectroscopy (TERS) is an optical imaging technique with a resolution far beyond the diffraction limit of light, which provides, simultaneously, scanning probe microscopy (SPM) and Raman spectroscopy information \cite{RN150, Stockle2000, Hartschuh2003, RN149, Hagen2005, MacIel2008, RN116, RN148, RN104, Yano2009, VanSchrojensteinLantman2012, Kumar2015, Park2016, Wang2017, Beams2018a}. It is based on the illumination of a sharp metallic tip that, on one hand, concentrates the incoming exciting electromagnetic field to a nanoscale near-field at the tip apex and, on the other hand, collects the near-field Raman scattering from the sample, resulting in a localized and enhanced stimulation of the sample's scattering \cite{RN112, RN94, Shi2017}. Therefore, it is not uncommon to simplistically assume that the TERS characteristics, including imaging resolution, is defined solely by the tip apex structure. However, several TERS experiments have now shown resolutions far beyond the tip apex dimension, achieving the nanometer scale in air \cite{RN134} and the angstrom scale in ultra-high vacuum \cite{RN100, Lee2019, Baumberg2019}. Such ``super-resolution" has been obtained using unexpected tricks, like local tip-induced pressure~\cite{Yano2009}, and in most cases utilizing the so-called gap mode configuration, where the enhancement can be further increased by locating the sample between the tip and a flat metallic substrate \cite{RN105,RN68, zhang2013chemical}. Whereas in the conventional TERS configuration the field enhancement at the tip apex is conventionally due to the excitation of localized surface plasmon resonance on the tip shaft~\cite{Vasconcelos2015, Vasconcelos2018}, the gap mode configuration makes use of the electric field enhancement by the gap-plasmon resonance that appears in the confined dielectric space between the tip end and the metallic substrate~\cite{Becker2016}.

In this work, we study the spatial distribution of the field enhancement during TERS experiments in three different TERS configurations: regular (non-gap mode), gap mode with a continuous metallic substrate, and a ``structured'' gap mode, utilizing regularly spaced metallic nanospheres as substrate. As a reference sample we utilize graphene, a strong and two-dimensional Raman scatterer~\cite{Jorio2011, Jorio2019}, which enables total surface sensing on top of the different substrates to show the significant influence of the substrate structure in the TERS results. We first introduce, in \cref{s:TD}, the technical aspects. In \cref{s:E} the experimental results are discussed, separated in three main findings: (A) TERS efficiency as a function of the tip-laser alignment within the focus; (B) TERS enhancement dependence on substrate structure and phonon symmetry; (C) the achievement of super-resolution for ``structured" gap mode. We then focus, in \cref{s:Tlta}, on how the ``structured'' gap mode is capable of generating an apparent super-resolution image. In~\cref{s:conc} we present the conclusions of this work.

\section{Technical Aspects}
\label{s:TD}

\subsection{Experimental Setup}
\label{s:exp_setup}

\begin{figure}[ht]
	\centering
	\includegraphics[width=8.5cm]{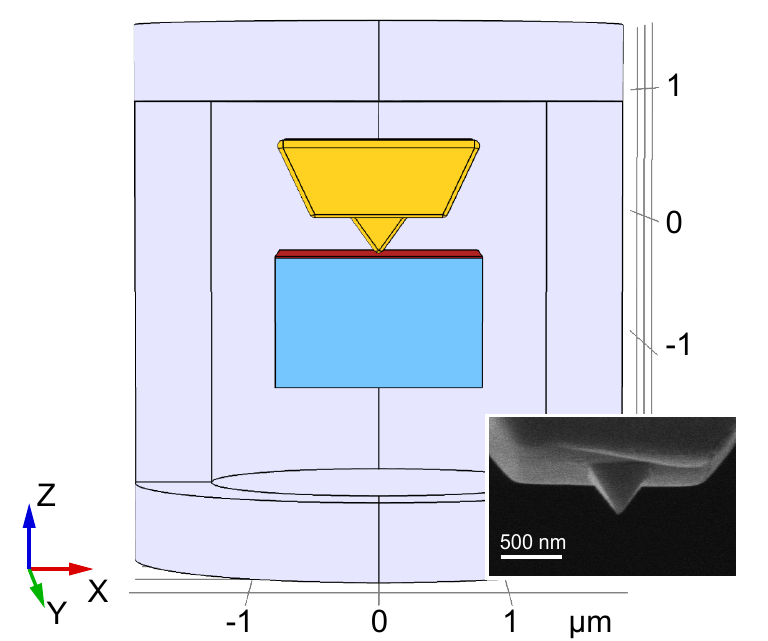}
	\caption{The simulation setup is enclosed by an air cylinder with $4.2\ \text{$\mu$m}$ height and $3.8\ \text{$\mu$m}$ diameter with $600\ \text{nm}$ thick PMLs at the simulation boundaries. The PTTP is shown in yellow, the thin film on top of substrate is shown in red and the glass substrate in light blue. The inset image portrays the SEM image of the actual tip used in some of the experiments.}\label{fig:simulation_setup}
\end{figure}

The TERS system consists of a combination of a non-contact atomic force microscope (AFM) and a micro-Raman spectrometer, optimized for a high numerical aperture (NA = 1.4) optical excitation and collection on a backscatter configuration~\cite{RN148}. The AFM setup is a home-made shear-force system with a tuning fork operating at $32.8\ \text{kHz}$, associated with a Phase-Locked Loop system that controls the tip-sample distance. Considering that the TERS setup is based on a radially polarized He-Ne laser beam with a 632.8 nm wavelength, a resonant gold pyramidal tip, denominated Plasmon-Tuned Tip Pyramid (PTTP), was used  (see inset to \cref{fig:simulation_setup}) \cite{Vasconcelos2018}. This tip is capable of holding localized surface plasmon resonance, in this case tuned for the given excitation wavelength. In addition, the tip used has a $40 \pm 10\ \text{nm}$ apex diameter, as measured by Scanning Electron Microscopy (SEM).

\subsection{Sample Preparation}
\label{s:TDE}

Graphene, as a spatially delocalized two-dimensional TERS sample, was prepared by the mechanical exfoliation method and deposited on three different substrates: (i) glass; (ii) a $12\ \text{nm}$ thick layer of gold evaporated on glass (Au film); (iii) a surface of $12\ \text{nm}$ diameter oleylamine-stabilized gold nanoparticles (AuNP) with an edge-to-edge inter-particle separation of $\approx 10\ \text{nm}$ between the particle surfaces in a roughly hexagonal lattice \cite{schulz2017size,mueller2018dark}.

\subsection{Group Theory Analysis}
\label{s:group_theory}

Considering graphene pertains to the $D_{6h}$ point group, the G band, observed at $\approx 1584$\ cm$^{-1}$ belongs to the $E_{2g}$ irreducible representation, while the second-order 2D band (also known as G' band), observed at $\approx 2700$\ cm$^{-1}$) is a totally symmetric $A_{1g}$ mode. The Raman tensors for the G and 2D bands of graphene, considering the presence of a highly focused field \cite{Budde2016}, are given by: 

\begin{equation}
    \alpha_G^1 = \begin{pmatrix} a & 0 & 0 \\ 0 & - a & 0 \\ 0 & 0 & 0 \end{pmatrix} , \qquad
    \alpha_G^2 = \begin{pmatrix} 0 & a & 0 \\ a &   0 & 0 \\ 0 & 0 & 0 \end{pmatrix} ,
    \label{eq:rotacoes3DG}
\end{equation}

and

\begin{equation}
    \alpha_{2D}^1 = \begin{pmatrix} b & 0 & 0 \\ 0 & b & 0 \\ 0 & 0 & c \end{pmatrix} , \qquad
    \alpha_{2D}^2 = \frac{1}{4}\begin{pmatrix} -b & -b\sqrt{3} & 0 \\ b\sqrt{3} & -b & 0 \\ 0 & 0 & c \end{pmatrix},
    \nonumber
\end{equation}
\begin{equation}
    \alpha_{2D}^3 = \frac{1}{4}\begin{pmatrix} -b & b\sqrt{3} & 0 \\ -b\sqrt{3} & -b & 0 \\ 0 & 0 & c \end{pmatrix}.
    \label{eq:rotacoes3D2D}
\end{equation}

From symmetry, the G band can only be activated by electric fields in the graphene (XY) plane. The 2D band can also be activated by fields polarized perpendicular to the graphene plane (Z). The $c$ value is not known in the literature, but experiments~\cite{Budde2016} indicate that $c \ll b$.

The selection rules for TERS have been derived by group theory~\cite{jorio2017symmetry}. The phonon active modes for the different scattering processes are defined by:

\begin{subequations}\label{eq_gt}
	\begin{gather}
	{\bf S}:(\Gamma_{vec} \otimes \Gamma_{vec})\subset\Gamma_{pn}\\
	{\bf SP}:(\Gamma_{vec} \otimes \Gamma^{{\cal H}_{pl-el}} \otimes \Gamma_{vec}) \subset\Gamma_{pn}\\
	{\bf PS}:(\Gamma^{{\cal H}_{pl-el}} \otimes \Gamma_{vec}) \subset\Gamma_{pn}\\
	{\bf PSP}:(\Gamma_{vec} \otimes \Gamma^{{\cal H}_{pl-el}} \otimes \Gamma^{{\cal H}_{pl-el}} \otimes \Gamma_{vec}) \subset\Gamma_{pn},
	\end{gather}
\end{subequations}
where {\bf S} is the usual Raman scattering Stokes process, where light interacts only with the sample; {\bf SP} and {\bf PS} are processes where the interaction of the incoming and outgoing light, respectively, is mediated by the plasmonic structure;
{\bf PSP} is a process where both incoming and outgoing light interactions are mediated by the plasmon. Notice Equation (3c) is different from what has been presented in Ref.\cite{jorio2017symmetry} because, in the case of a radially polarized incoming excitation (as utilized in the experimental setup described in \cref{s:exp_setup}), the PS light-induced excitation of the plasmonic tip occurs via a totally symmetric field distribution rather than a vector-like linearly polarized excitation. 

The difference on going from non-gap mode to gap mode TERS is that the TERS system changes from the $C_{\infty v}$ point group to the $D_{\infty h}$ due to the mirror symmetry imposed by the metallic surface. When comparing regular TERS ($C_{\infty v}$) with gap mode TERS ($D_{\infty h}$), the PS scattering becomes forbidden for both the G and 2D bands in gap mode. 

\subsection{Frequency Domain Simulations for Far-Field and Near-Field Distributions}
\label{s:TDT}

The simulations are based on the experimental setup described in \cref{s:exp_setup}. \Cref{fig:simulation_setup} displays the positioning of tip and substrate in the simulation environment. The simulations were performed using the Finite Element Method (FEM) implemented by the Comsol Multiphysics V in the frequency domain. The tip utilized in the simulations was a PTTP tuned for a $632.8\ \text{nm}$ excitation wavelength with an apex diameter of $40\ \text{nm}$ and an internal angle of $70.54^{\circ}$ between pyramid faces. The boundaries are treated with a $600\ \text{nm}$ thick Perfectly Matched Layer (PML). All the components not composed of air are not in contact with the PML to avoid calculation artifacts. The tip-sample gap is set to $5\ \text{nm}$ for all cases, as to properly simulate the gap for non-contact AFM and to take advantage of the light confinement~\cite{Novotny2012}. The gold material model utilized for the PTTP tip, the gold film and the AuNP were obtained experimentally from reflection and transmission measurements of thin gold films by Johnson and Christy~\cite{Johnson1972}.

As for the input electromagnetic field, a radially polarized, tightly focused Gaussian beam was modeled using the paraxial approximation for a Gaussian beam with $360\ \text{nm}$ waist diameter and polarization along the vertical axis (direction of propagation). Since the excitation is purely polarized on the vertical axis, the Gaussian beam waist diameter accounts only for the central Z lobe size in a system with a $1.4$ numerical aperture (provided by an oil immersion objective lens) and a $632.8\ \text{nm}$ excitation wavelength~\cite{Novotny2012}.

In order to reduce computational costs, the simulation environment was truncated at symmetry planes corresponding to $x = 0\ \text{nm}$ and $y = 0\ \text{nm}$, resulting in a quarter section of the original environment. The resulting new boundaries were treated as perfect magnetic conducting surfaces in order to impose symmetry to the electric field with respect to the cut planes.

\begin{figure}[ht]
	\centering
	\includegraphics[width=8.5cm]{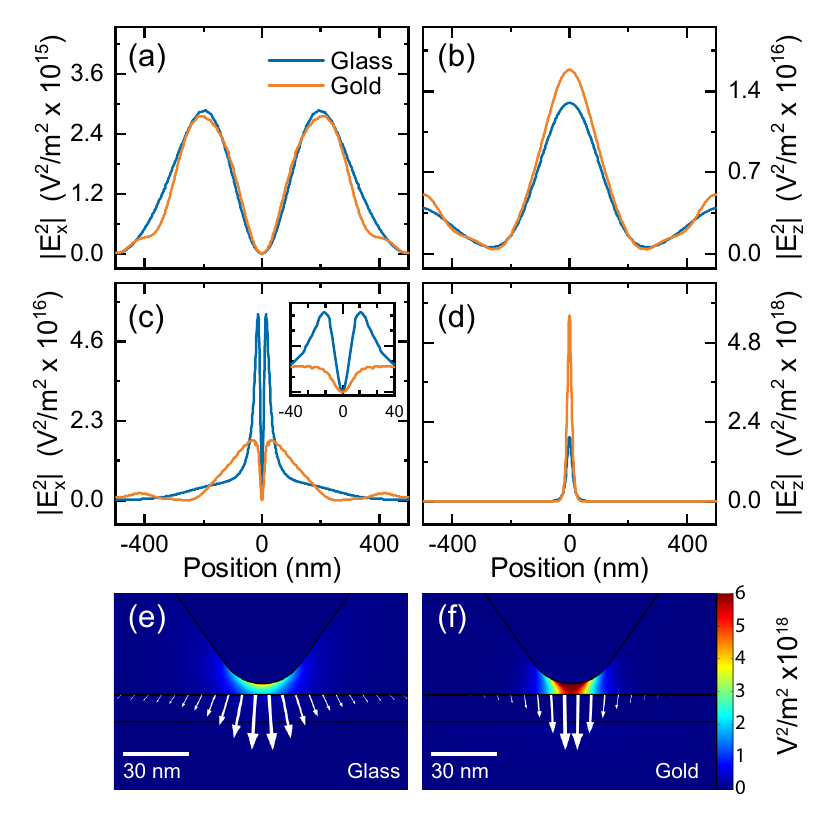}
	\caption{Simulations of the electric field at the sample plane for glass and gold substrates. (a,b) show the intensity profile of the $|E_x^2|$ and $|E_z^2|$, respectively, in the absence of the tip; (c,d) display the field intensity at the same location, but with the tip at $5\ \text{nm}$ distance from the substrate. The inset to (c) is a zoom up close to the tip location ($0\ \text{nm}$ position); (e) shows the color coded intensity map with electric field orientation at the glass substrate portrayed by white arrows and (f) is the equivalent result for the $12\ \text{nm}$ gold film substrate.}\label{fig:confocalTERS}
\end{figure}

\Cref{fig:confocalTERS} (a,b) describes the far-field (no tip) and (c--f) near-field (with tip) intensity distributions obtained by the frequency-domain modeling, considering the outlined specifics of our experimental setup. The distinction between the field intensity distribution in the presence of glass or gold substrate is obtained, where the blue curves stand for non-gap mode and the orange curves stand for the gap mode configurations in~\cref{fig:confocalTERS}(a--d). The left (a, c) and right (b,d) panels stand for the electric field polarization parallel (X, in-plane) and perpendicular (Z, out-of-plane) to the substrate plane, respectively. In~\cref{fig:confocalTERS}(e,f) the field vectors at the sample's plane are displayed as white arrows.

Finally, we also performed two-dimensional simulations to understand super-resolution results obtained on the structured gap-mode configuration, where the computational costs get too high due to the loss of the square-lattice plasmonic symmetry. Further details on \cref{s:Tlta}.

\section{Experimental Results and Discussions}
\label{s:E}

\subsection{Tip Scanning the Diffraction Limited Confocal Illumination Area}
\label{s:Ets}

\begin{figure}[ht]
	\centering
	\includegraphics[width=8.5cm]{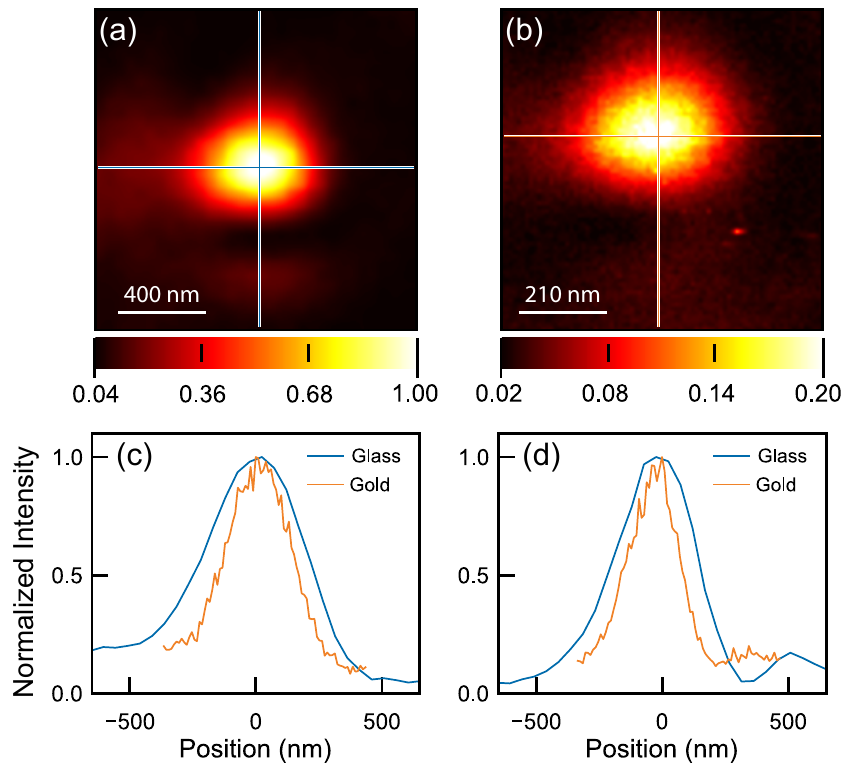}
	\caption{(a) Normalized 2D band intensity map for graphene on glass generated by scanning the tip within the diffraction limited laser focus; (b) Equivalent map for gold film substrate (generated with the same tip), normalized by the same value as in (a); (c,d) Horizontal and vertical line profiles, respectively, for gold and glass substrates, from the locations indicated with corresponding colors in (a, b), but here with line profiles normalized to 1.}\label{fig:sample_field_gold_glass}
\end{figure}

In conventional TERS setups the AFM gold tip is aligned and fixed with respect to the laser focus, and the sample is moved along the XY plane by a piezo stage. In order to study the TERS spatial distribution around the laser focus, we scanned the focal region in the XY plane by moving the tip with respect to the fixed laser spot (and sample), measuring the Raman signal intensity of graphene's 2D band. This procedure was made for the graphene on glass (non-gap mode) and for the graphene on top of the thin gold film (gap mode). By plotting the 2D band intensity as a function of tip position, we identify the spatial distribution of the convolution between near-field tip response and laser spot, as shown in~\cref{fig:sample_field_gold_glass}(a,b). The maximum 2D band TERS intensity is obtained in the central ($0\ \text{nm}$) position in both configurations. The full-width at half maximum (FWHM) is smaller in the gap mode configuration: $429\ \text{nm}$ for the glass and $291\ \text{nm}$ for gold, a $32\%$ reduction for the gap mode configuration. Similar (although less intense) results are observed for the G band TERS.

The sharper TERS distribution for gap mode can be understood based on the field distributions shown in \cref{fig:confocalTERS}. For the far-field (\cref{fig:confocalTERS}(a,b)) there is a difference in the spread of the in-plane X-polarized field, which is slightly more compressed towards the center, accompanied by a small increase in the very central intensity of the out-of-plane Z-polarized field. For the near-field configuration (\cref{fig:confocalTERS}(c,d)) the difference also depends on the direction of the electric field. For Z-polarized near-field (d), the field distribution in gap mode is $70 \%$ more intense and $26\%$ narrower than for the non-gap mode. For the X-polarized near-field (c), however, it is the opposite when looking closer to the central area under the tip, and the trend in the most intense signal exhibits inversions on each configuration (non-gap mode vs. gap mode, blue and red curves, respectively) as the displacement from the central position increases. Overall, there is a sharper field distribution for the gap mode. The differences in TERS localization are even stronger considering that TERS intensity is proportional to electric field powers up to $|E|^4$ \ \cite{Yang2009}. It is important to note, however, that, for this analysis, care has to be taken in proper alignment, since a change (maybe due to experimental drift) in the focus condition between these two experiments can also cause changes in the FWHM.

\subsection{Near- and Far-Field Comparison for Different Symmetry Modes and Different Substrates}
\label{s:Etutd}

\begin{figure}[ht]
	\centering
	\includegraphics[width=8.5cm]{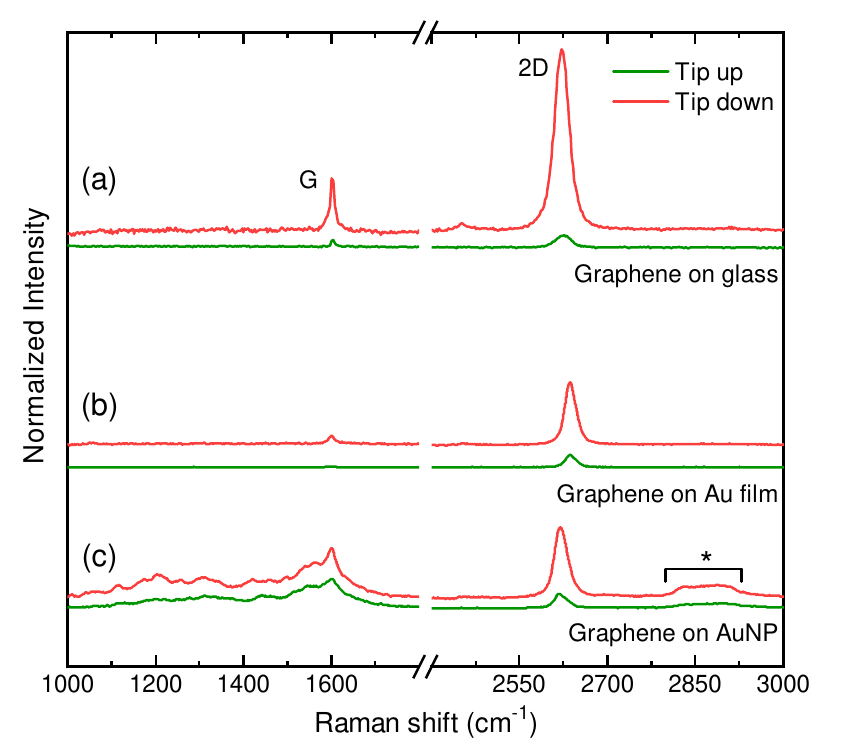}
	\caption{Raman spectra of graphene placed on three different substrates ((a) glass; (b) Au film; (c) AuNP) in the spectral range of the G and 2D bands. Tip down and tip up spectra are marked red and green, respectively. All spectra are normalized to exhibit the same normalized 2D band amplitude on tip up condition an were acquired with the same excitation power. * Indicates the oleylamine feature utilized to plot \cref{fig:ivsd}(a, b).}\label{fig:tutd}
\end{figure}

We now analyze how different substrates influence the total spectral enhancement when the tip is placed in the optimal location for TERS signal, i.e. at position $0\ \text{nm}$ in~\cref{fig:sample_field_gold_glass}(c,d). The spectral enhancement factor is defined here as $F_{\text{TERS}}=A_{\text{NF}}/A_{\text{FF}}$, where $A_{\text{NF}}$ is the integrated intensity (area) of a Raman peak in the presence of the tip (NF standing for near-field) and $A_{\text{FF}}$ the equivalent value in the same region with the tip retracted far away from the sample (FF standing for far-field).~\Cref{fig:tutd} shows the graphene Raman spectra with and without the tip on the three different substrates.

The enhancement factors $F_{\text{TERS}}$ were measured for the G ($E_{2g}$) and 2D ($E_{1g}$) bands on glass, Au film and AuNP substrates, resulting on the values summarized in~\cref{tab:enhacements}.

\begin{table}[ht]
	\centering
	\caption{G and 2D band enhancement factors $F_{\text{TERS}}$ for the three  substrate types.}\label{tab:enh}
	\begin{tabular}{c|c|c|c|}
		                         \cline{2-4} & Glass & Au film & AuNP \\ \hline
		\multicolumn{1}{|l|}{\textbf{G}}     & $10 \pm 4$  & $7 \pm 1$   & $5 \pm 3$  \\ \hline
		\multicolumn{1}{|l|}{\textbf{2D}}    & $16 \pm 1$  & $5 \pm 1$   & $5 \pm 1$  \\ \hline
	\end{tabular}\label{tab:enhacements}
\end{table}

The average results and the uncertainties were obtained analyzing seven tip up and seven tip down spectra like the ones shown in~\cref{fig:tutd}, obtained during a scanning procedure of homogeneous regions (accumulation time of 2 seconds per point for graphene on glass, 5 seconds for graphene on Au film and 10 seconds for graphene on AuNP, excitation power of $160\ \text{$\mu$W}$ at the sample for all cases). The estimated uncertainty is larger for the G band on the AuNPs substrate because of the presence of the oleylamine peaks in the second case (see~\cref{fig:tutd}(c), near 1600\ cm$^{-1}$). Interestingly, the enhancement factors change depending on the substrate and the Raman band.
Counter-intuitively, the overall enhancement is larger for regular TERS (on glass) as compared to the gap mode configurations, consistent with what has been shown in \cref{fig:sample_field_gold_glass}(a, b). 

This counterintuitive result can be understood based on the group theory analysis, combined with the electric field distributions shown in~\cref{fig:confocalTERS}. The presence of the conductive substrate strongly enhances the Z-polarized field, but not the XY-polarized fields. Since graphene responds to electric fields along the plane, the gap mode is actually not effective in enhancing the Raman response of this two-dimensional system. It is important to note that this result implies that the out-of-plane response of totally symmetric (2D) mode, although not symmetry forbidden, is truly negligible, i.e. the Raman tensor parameter $c \approx 0$ (see \cref{eq:rotacoes3D2D}). As it can be seen in~\cref{tab:enh}, while on glass the enhancement factor of the 2D band intensity is 16, on gold it is roughly three times smaller.  

Besides, when comparing the results obtained for the G ($E_{2g}$) and 2D ($A_{1g}$) modes, the result on glass is consistent with reports on the literature, where the 2D band enhances more than the G band due to near-field coherence effects that privilege totally symmetric modes \cite{beams2014spatial,canccado2014theory}. Interestingly, this difference washes out in the gap mode configuration, and again, this can be understood as due to the stronger confinement of the field very near the tip location. The inset to \cref{fig:confocalTERS}(c) shows that, in the gap mode, the in-plane field is strongly reduced close to the tip location, within the phonon coherence length ($\sim 30\text{nm}$), where the near-field interference effects take place~\cite{beams2014spatial}. This also confirms that the higher enhancement for the 2D band on glass is due to the non-local PS and SP scattering~\cite{jorio2017symmetry, canccado2014theory}, rather than due to the out-of-plane $c$ component of the Raman tensor, otherwise the 2D band should enhance more (not less) in gap mode (see~\cref{tab:enh}). 

\subsection{TERS Line Profile in Structured gap mode}
\label{s:Lsgm}

\begin{figure}[ht]
	\centering
	\includegraphics[width=8.5cm]{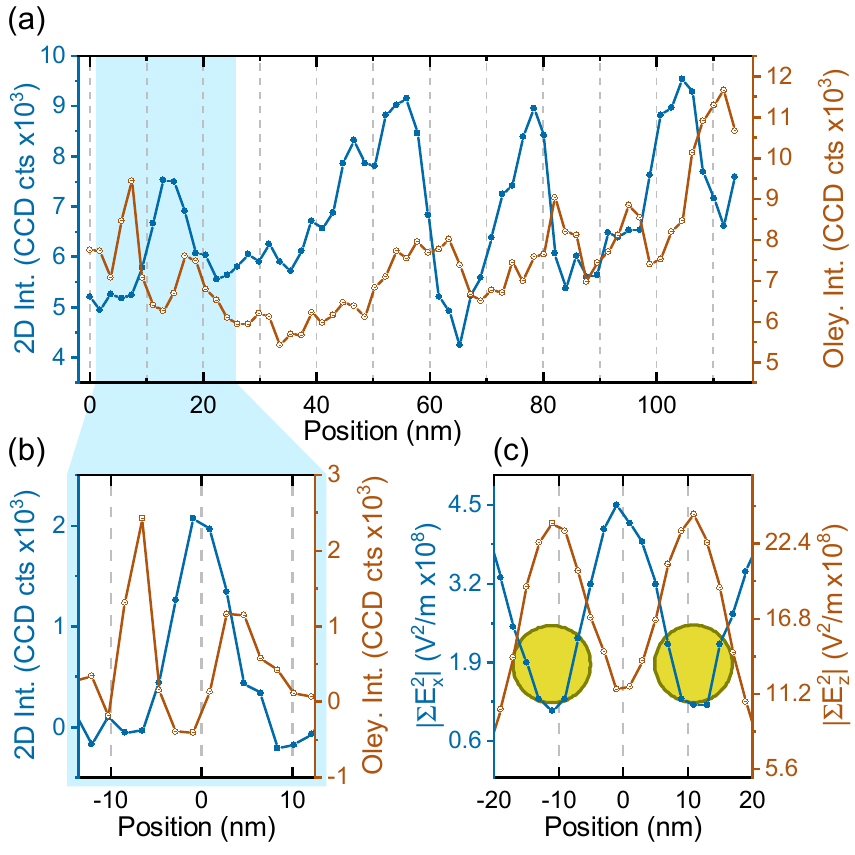}
	\caption{(a) Intensity profile of the graphene 2D (blue filled bullets) and the oleylamine (orange open bullets) Raman peaks when moving the TERS tip along a line scan over AuNP (excitation power of $160\ \text{$\mu$W}$ and integration time of $2.5\ \text{s}$ per point); (b) Detail of section highlighted in (a) with linear backgrounds removed to improve visualization; (c) Simulated intensities of integrated in-plane (X, blue) and out-of-plane (Z, orange) electric field components along a line scan. The golden circles indicate the positions and sizes of the AuNP in the simulation.}
	\label{fig:ivsd}
\end{figure}

To test field localization and the possible achievement of ultra-high resolution, we measured graphene on top of the AuNP substrate, as described in~\cref{s:TDE}, while scanning the substrate. Since graphene is homogeneously present in this sample, the only variation throughout the scan is the configuration of AuNP underneath the tip's apex.~\Cref{fig:ivsd}(a) shows the intensity trends of the 2D phonon mode (blue filled bullets) and also the intensity of a nearby Raman band (orange open bullets, $\approx 2850$\ cm$^{-1}$, see * in \cref{fig:tutd}(c)), attributed to oleylamine, during the line scan. Since the AuNPs are coated by  a layer of oleylamine required for the self-assembly into an AuNP monolayer, its Raman band can also be observed.

Considering that the tip used for the experiment shown in \cref{fig:ivsd} has a $~40\ \text{nm}$ diameter, the total scan of $120\ \text{nm}$ is a relatively small scanning region. Still, clear oscillations in the 2D and oleylamine Raman intensities are observed. In terms of lateral resolution, a Fast Fourier Transform analysis of the 2D band intensity map from which the line profile in~\cref{fig:ivsd}(a) was taken, results in a spatial resolution of $6.7\ \text{nm}$, close to the Nyquist limit of $3.75\ \text{nm}$ expected for the $1.875\ \text{nm}$ per pixel sampling rate utilized. This can be considered super-resolution given the $40\pm 10 \ \text{nm}$ apex diameter for the tip utilized in this experiment. 

Interestingly, we observe that whenever the intensity of the 2D band increases, the intensity of the oleylamine band decreases. The alternating peak intensity locations when comparing the 2D band and oleylamine bands can be explained considering the intensity profile trends of the in-plane (X) and out-of-plane (Z) components of the electric field as the sample is scanned, as shown in the simulation results in~\cref{fig:ivsd}(c) (more detail in \cref{s:Tlta}). Note the similarity between the simulation (\cref{fig:ivsd}(c)) and the detailed experimental section in \cref{fig:ivsd}(b). The 2D band is maximum when the in-plane X field is maximum, which happens between particles, while the oleylamine peaks are maximum when the out-of-plane Z field is maximum, which happens on top of a particle.

\section{Further Simulations and Discussions on super-resolution}
\label{s:Tlta}

\Cref{s:Lsgm} showcases how increased resolution can be obtained from a structured gap mode substrate. In this section, frequency domain simulations using an adapted version of the setup described in~\cref{s:TDT}, limited to two dimensions are used in order to properly characterize the field distribution when ``super-resolution" situation is achieved. In the simulation environment, the substrate is modeled by 50 gold circles with a diameter of $12\ \text{nm}$ and a $10\ \text{nm}$ gap between each other.

\begin{figure}
	\includegraphics[width= 8.5cm]{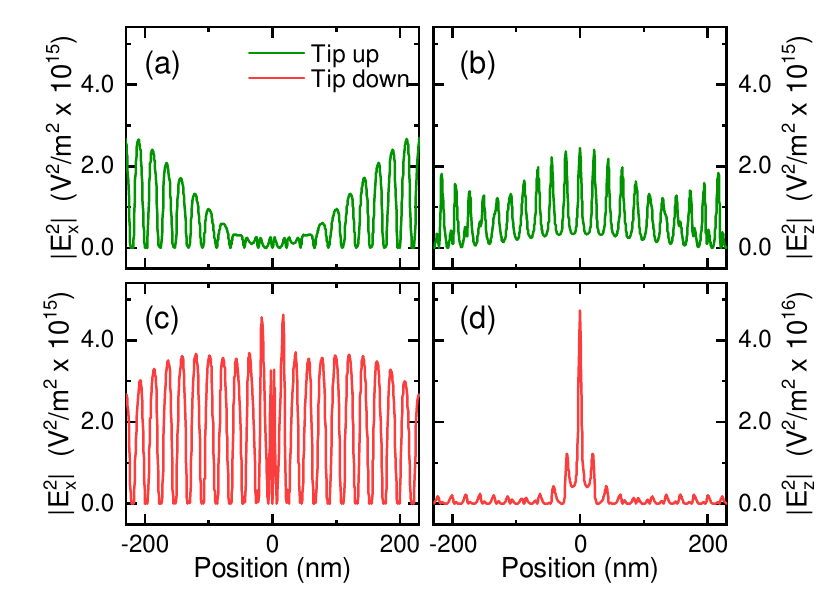}
	\caption{Simulated tip up (a, b) and down (c, d) distributions for in-plane (a, c) and out-of-plane (b, d) squared electric field components. To probe the field distribution, the values of the electric field were plotted over a horizontal line $4.9\ \text{nm}$ away from the tip's apex and $0.1\ \text{nm}$ away the nanospheres, which is roughly were the graphene would be located.} 
	\label{fig:raw_fields_tip_up_down}
\end{figure}

\Cref{fig:raw_fields_tip_up_down} shows a simulated tip up and down experiment. For this structured substrate, the profiles observed in \cref{fig:confocalTERS}(a-d) are now superposed by modulations induced by the AuNP. When the tip is landed (red traces), there is an increase in field intensity. However, the enhancement is localized near the tip for the out-of-plane Z-field component (\cref{fig:raw_fields_tip_up_down}(d)), but completely delocalized for the in-plane X-field component (\cref{fig:raw_fields_tip_up_down}(c)). Therefore, for the experiment shown in \cref{fig:ivsd}(a,b), while the oleylamine spectra comes majorly from molecules localized under the tip, the picture is completely different for the graphene 2D band. 

\begin{figure}
	\includegraphics[width= 8.5cm]{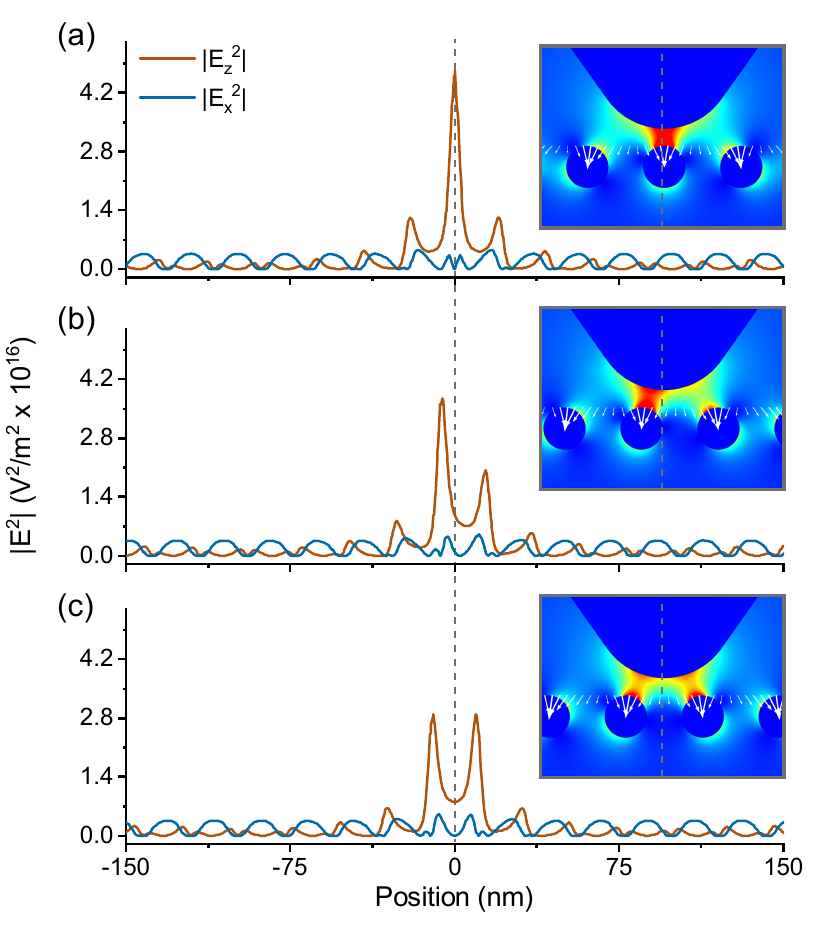}
	\caption{Electric field intensity distribution for horizontal and vertical components for distinct particle configurations under the tip: (a) tip on top of a particle, (b) slightly misaligned with a particle and (c) in between particles. The fields are plotted over the same region described in \cref{fig:raw_fields_tip_up_down}. The insets to each graph portray a color coded 2D distribution of $|E|^2$.}
	\label{fig:aunp_raw_field}
\end{figure}

For further details, \cref{fig:aunp_raw_field} shows the changes in the X- and Z-field intensities for different relative position of the tip with respect to the AuNPs, i.e. right on top of a particle (a), exactly in between two particles (c), and between these two cases (b). In all cases, the graphene TERS signal (given by the X-polarized field) should come from the entire focal region, while some degree of localization is only obtained for TERS related to the Z-polarized field component.

\section{Conclusions}
\label{s:conc}

By exploring experimentally and theoretically the TERS electric field distribution in graphene on different substrates, we consistently found that in gap mode configuration a strong Z-polarized field is excited, but it does not generate extra enhancement for 2D systems such as graphene, which responds to electric fields polarized along the substrate plane. Our analysis solidifies the conclusion that the totally symmetric modes in graphene have a negligible Raman response for fields polarized perpendicular to the graphene plane, even if not symmetry forbidden. Furthermore, we show that near-field interference effects are suppressed for the in-plane fields in gap mode.

Additionally, it was shown, both by simulations and experiments, that the composition of the substrate has an effect on field confinement and, consequently, on resolution. Nevertheless, the resolution can be further improved, beyond the tip's apex diameter, by means of a careful choice of the tip-sample-substrate interaction. For instance, a conductive substrate with features smaller than the tip's apex, such as gold nanoparticles, can be employed to improve the lateral resolution, but this is effective only for the out-of-plane polarized field. However, the substrate actually delocalizes the in-plane electric field. This effect must be carefully considered when analysing sub-nanometer TERS measurements, as the tip will still interact with all sub-nanometer features in its vicinity. Although our results were developed for nanometer-size structures, similar effects should be observed in pico-cavity measurements~\cite{Lee2019, Baumberg2019}.

\section{Acknowledgements}
This work was supported by the Funda\c{c}\~{a}o Coordena\c{c}\~{a}o de Aperfei\c{c}oamento de Pessoal de N\'\i vel (CAPES) and the Deutsche Akademische Austauschdienst (DAAD) within the PROBRAL program under grant number 57446501. A.J. acknowledges financial support from the Humboldt Foundation and CNPq (552124/2011-7, 307481/2013-1, 304869/2014-7, 460045/2014-8, 305384/2015-5, 309861/2015-2. H.M. and C.R. acknowledges financial support from Finep, CNPq and Fapemig. P.K. and S.R. acknowledge support by the European Research Council ERC under grant DarkSERS (772108) and the Focus Area NanoScale of Freie Universit\"at.

\end{document}